\documentclass[prd,twocolumn,floatfix,preprintnumbers,nofootinbib,nopacs]{revtex4}
\usepackage[utf8x]{inputenc}
\usepackage{mathrsfs}
\usepackage{amssymb}
\usepackage{amsmath}
\usepackage{graphicx}

\usepackage[caption = false]{subfig}
\usepackage[normalem]{ulem}
\usepackage{xcolor}
\definecolor{lcolor}{rgb}{0.5,0,0}
\definecolor{citcolor}{rgb}{0,0.3,0.0}
\usepackage{nicefrac}

\usepackage[breaklinks,colorlinks,urlcolor=blue,citecolor=citcolor,linkcolor=lcolor]{hyperref}
\usepackage{mciteplus}

\usepackage{physics}

\newcommand{\rt}{{\mathbf{r}}}
\newcommand{\xt}{{\mathbf{x}}}
\newcommand{\bt}{{\mathbf{b}}}

\newcommand{\yt}{{\mathbf{y}}}

\newcommand{\nc}{{N_\mathrm{c}}}

\newcommand{\as}{\alpha_{\mathrm{s}}}

\newcommand{\xT}{\mathbf{x_\perp}}

\begin{document}
\author{Heikki Mäntysaari}
\email{heikki.mantysaari@jyu.fi}
\author{Pragya Singh}
\affiliation{
Department of Physics, University of Jyväskylä,  P.O. Box 35, 40014 University of Jyväskylä, Finland
}
\affiliation{
Helsinki Institute of Physics, P.O. Box 64, 00014 University of Helsinki, Finland
}

\begin{abstract}
    We quantify the  effect of high-energy JIMWLK evolution on the deformed structure or heavy (Uranium) and intermediate (Ruthenium) nuclei. The soft gluon emissions in the high-energy evolution are found to drive the initially deformed nuclei towards a more spherical shape, although the evolution is slow ,especially for the longest distance-scale quadrupole deformation.   We confirm a linear relationship between the squared eccentricity $\varepsilon_n^2$ and the deformation parameter $\beta_n^2$ in central collisions across the energy range covered by the  RHIC and LHC measurements. The applied JIMWLK evolution is found to leave visible signatures in the eccentricity evolution that can be observed if the same nuclei can be collided at RHIC and at the LHC, or in rapidity-dependent flow measurements. Our results demonstrate the importance of including the Bjorken-$x$ dependent nuclear geometry when comparing simulations of the Quark Gluon Plasma evolution with precise flow measurements at high collision energies.
\end{abstract}

\title{Energy dependence of the deformed nuclear structure at small-$x$}

\maketitle

\section{Introduction}
Understanding the high-energy structure of protons and nuclei is of fundamental interest. This structure is a result of complex interplay between different QCD phenomena.
At the the confinement scale, strong interactions are responsible for forming a color-neutral bound state out of quarks and gluons. 
At high collision energies, this structure is dominated by a large number of soft gluons emitted throughout the high-energy QCD evolution, which itself can be understood by weak coupling techniques. 

A simple pointlike probe that can be used to probe the internal structure of protons and nuclei is a virtual photon, emitted by a lepton in Deep Inelastic Scattering (DIS). Thanks to the precise electron-proton DIS measurements performed at HERA~\cite{H1:2015ubc}, the partonic content of the proton, as well as the proton geometry and how it depends on Bjorken-$x$, is relatively well known. In the next decade, the Electron-Ion Collider in the US~\cite{AbdulKhalek:2021gbh} will enable similar studies with nuclear targets, for the first time in collider kinematics. One particular advantage of the EIC is its capability to accelerate almost any nuclei, making it possible to also study specific nuclei with interesting geometric structures.

Nuclear geometry is also a crucial input for the studies of Quark Gluon Plasma (QGP) in relativistic heavy ion collisions. This is because the hydrodynamical evolution of the Quark Gluon Plasma (QGP) will transform initial state pressure (density) anisotropies into observable final state correlations~\cite{Gardim:2011xv}. 
Detailed knowledge of the initial nuclear geometry is especially important when studying collisions among deformed nuclei~\cite{Filip:2009zz,Masui:2009qk,Hirano:2012kj,Shen:2014vra,Schenke:2014tga,Mantysaari:2017cni,Schenke:2020mbo,Xu:2021uar,Giacalone:2021udy,Zhang:2021kxj,Nijs:2021kvn,Zhao:2022mce,Bally:2022vgo,Bally:2023dxi}, such as U+U, Ru+Ru or Zr+Zr collisions at RHIC~\cite{STAR:2021mii,STAR:2015mki} or Xe+Xe collisions at the LHC~\cite{ALICE:2024nqd}. 
These experiments provide unique opportunities to probe the properties of the Quark Gluon Plasma in systems with vastly different geometries. Additionally, they bridge the gap between the low-energy and high-energy nuclear physics communities. For example, it was shown in Ref.~\cite{Ryssens:2023fkv} that the U+U collisions measured at RHIC provide evidence for the hexadecapole deformation of the ${}^{238}$U nucleus. Further possibilities to probe the nuclear structure in high-energy nucleus-nucleus collisions have been discussed e.g. in Refs.~\cite{Singh:2023rkg,Giacalone:2024ixe,Mantysaari:2024uwn}. On the other hand, in Ref.~\cite{Mantysaari:2023qsq} it was demonstrated that at the future EIC, diffractive processes in electron-nucleus scattering can probe quadrupole, octupole and hexadecapole deformations of the heavy nuclei. Before the EIC is realized, similar studies are also possible in ultra-peripheral collisions~\cite{Klein:2019qfb} as demonstrated e.g. in Refs.~\cite{Lin:2024mnj,Mantysaari:2023prg}.

The purpose of this work is to quantify how the deformed nuclear geometry changes with the collision energy, or with the rapidity at which the measurement is performed. This evolution we calculate within the Color Glass Condensate (CGC)~\cite{Iancu:2003xm} framework which provides a convenient method to describe QCD in the high energy (high parton density) domain. Within the CGC, the energy (or equivalently rapidity) dependence of the nuclear structure is obtained
by solving the JIMWLK~\cite{Jalilian-Marian:1997qno,Jalilian-Marian:1997ubg,Kovner:2000pt,Iancu:2000hn,Mueller:2001uk} high-energy evolution equation. 
Furthermore,  using eccentricities as a proxy for flow measurements, we quantify the expected effects of the JIMWLK evolution on the collectivity measurements in nucleus-nucleus collisions. We argue that a proper interpretation of the flow measurements probing the Quark Gluon Plasma properties requires a detailed understanding of the energy-dependent nuclear geometry.

\section{Energy dependent nuclear geometry}
\label{sec:setup}
At high energies where parton densities are very large, it is convenient to describe the nucleus as a dense color field. The nuclear structure can be probed by a simple perturbatively understandable probe, which we take to be a small $q\bar q$ dipole. As such, this corresponds to a DIS process in the dipole picture where the photon splits to a dipole long before the scattering. Unlike in an actual experiment, for the purposes of this work we consider a fixed-size dipole and fix the orientation of the deformed nucleus in order to probe (or define) the energy-dependent nuclear density.

The dipole-nucleus scattering we calculate within the CGC approach. The nuclear (deformed) high-energy structure at the initial condition of the small Bjorken-$x$ evolution, taken to be $x=x_0=0.01$, is described as in the IP-Glasma~\cite{Schenke:2012wb} framework applied recently in a similar context e.g. in Refs.~\cite{Singh:2023rkg,Mantysaari:2022sux,Mantysaari:2023qsq,Mantysaari:2024uwn,Mantysaari:2023prg}. We first sample the nucleon positions from the Woods-Saxon distribution
\begin{equation}\label{eq:WS}
    \rho(r,\theta) = \frac{\rho_0}{1+\exp[(r-R(\theta))/a]}\,,
\end{equation}
where the effective radius depends on the azimuthal angle as
\begin{equation}
\label{eq:deform_R}
    R(\theta)=R[1+\beta_2 Y_2^0(\theta)+\beta_3 Y_3^0(\theta) +\beta_4 Y_4^0(\theta)].
\end{equation}
Here we have oriented our coordinate system such that, in the case of quadrupole deformation $\beta_2>0$, the long axis corresponds to $\theta=0$. The skin depth $a$ is nucleus-dependent, and taken to be independent of $x$ in this work. The transverse thickness function $T_A(\bt)$ is defined as
\begin{equation}
\label{eq:TA}
    T_A(\bt) = \int \dd{z} \rho\left(\sqrt{\bt^2+z^2}\right).
\end{equation}

The small-$x$ color field of the nucleus is described using the impact parameter dependent McLerran-Venugopalan (MV) model~\cite{McLerran:1993ka,McLerran:1993ni}. Within the MV model, the color charge is assumed to be a  local random Gaussian variable with a correlator 
\begin{multline}
\label{eq:mv_rhorho}
    g^2 \langle \rho^a(x^-,\xt) \rho^b(y^-,\yt)\rangle = \delta^{ab} \delta^{(2)}(\xt-\yt) \delta(x^- - y^-) \\
    \times g^4 \lambda_A(x^-,\xt).
\end{multline}
Here the local color charge density $\mu^2(\xt) = \int \dd{x^-} \lambda_A(x^-,\xt)$  is related to the saturation scale at point $\xt$, extracted from the IPsat parametrization~\cite{Kowalski:2003hm,Rezaeian:2012ji}. Moreover, the saturation scale is proportional to the local thickness function $T(\bt)$ obtained by integrating over the $z$ axis, which we obtain by summing over individual nucleon density profiles that are assumed to be Gaussian and have a width $B= 4$ GeV$^{-2}$. Once the color charge distribution $\rho(x^-,\xt)$ is known, one can solve for the Wilson lines that describe the eikonal propagation of a quark within the target color field:
  \begin{equation}
    V_{\xt} = P_- \exp \left\{ -ig \int \dd{x^-} \frac{\rho^a(x^-,\xt) t^a}{\nabla^2_{\xt} - \tilde m^2} \right\}.
\end{equation}
Here an infrared regulator $\tilde m=0.2~$GeV  is included in order to regulate the Coulomb tails. The dipole-nucleus scattering amplitude can then be written in terms of the Wilson lines as 
\begin{equation}
    N(\rt,\bt;Y)  \label{eq:dipoleamp} 
     = \left\langle 1 - \frac{1}{\nc} \tr \left[ V\left(\bt + \frac{\rt}{2}\right) V^\dagger\left(\bt - \frac{\rt}{2}\right) \right]\right\rangle . 
\end{equation}
Here the Wilson lines $V(\xt)$ (in the fundamental representation) depend implicity on the evolution rapidity $Y=\ln\frac{x_0}{x}$. The average $\langle \mathcal{O}\rangle$ refers to the average over possible target configurations.

The  Bjorken-$x$ dependence of the Wilson lines is obtained by numerically solving the JIMWLK evolution equation~\cite{Cali:2021tsh} (see also Ref.~\cite{McDonald:2023qwc} for a recent development on including the JIMWLK evolution in the IP-Glasma). 
To regulate Coulomb tails in the evolution, the evolution kernel describing the gluon emission is modified at large distances following Refs.~\cite{Schlichting:2014ipa,Mantysaari:2018zdd} as
\begin{equation}
\label{eq:jimwlk_kernel}
    K^i_{\xt} = \frac{x^i}{\xt^2} \to m |\xt| K_1(m|\xt|) \frac{x^i}{\xt^2}.
\end{equation}
Here we use $m=0.2$ GeV as a regulator and a fixed coupling $\as=0.15$ in the JIMWLK evolution. In this work, we do not aim to determine optimal values for the model parameters (see however related works~\cite{Mantysaari:2018zdd,Mantysaari:2022sux}), but use these phenomenologically motivated values. The conclusions of this work can be expected to be independent on the actual values chosen for these parameters. Generically running coupling effects should result in faster evolution at large distance scales, and as such the running coupling corrections can be expected to enhance the energy-dependence of the deformed geometry obtained in this work.

Although the dipole-nucleus amplitude is a clean probe of the nuclear high-energy structure, and we can use it to characterize the local density, its impact parameter dependence is not directly observable\footnote{We however note that that the impact parameter dependence is directly related to the momentum transfer dependence in exclusive processes~\cite{Klein:2019qfb,STAR:2017enh}}. More importantly, in an actual DIS process, it is not possible to control the orientation of the deformed nucleus. As such, our strategy in this work is to first use the dipole-nucleus scattering to extract the JIMWLK-evolved nuclear geometry.  Then we study eccentricities in nucleus-nucleus collisions. As eccentricities are strongly correlated with the measurable final state particle flow harmonics~\cite{Gardim:2011xv}, this provides us a method to determine the sensitivity of the flow measurements on the energy-dependent nuclear geometry. 
As we will demonstrate in this work,  there is a clean correlation between the eccentricities and the extracted deformation parameters, demonstrating that eccentricity (flow) measurements can probe the $x$-dependent deformed nuclear structure.

Eccentricities directly reflect the shape asymmetries in the initial-state geometry of heavy-ion collisions \cite{Rosenhauer:1986tn,Gupta:2000si,Zhang:2021kxj,Goldschmidt:2015kpa,Giacalone:2018apa}. We adopt the standard approach to characterize the event geometry using eccentricities, defined as:
\begin{align}\label{eq:ecc}
        \varepsilon_n(Y)=\frac{\int \dd[2]{\xt} T^{\tau\tau}(Y,\xt)~|\xt|^n e^{in\phi_{\xt}}}{\int \dd[2]{\xt} T^{\tau\tau}(Y,\xt)~|\xt|^n}\,,
\end{align}
where $\phi_\xt$ is the azimuthal angle of the vector $\xt$ and $T^{\tau \tau}$ is the energy density. The energy density is obtained as follows. Immediately after the collision (at $\tau=0^+$) in the forward light cone the non-zero components of the gauge fields read
\begin{eqnarray}
&&A^{i}_{\xT}(\tau=0^{+})=\frac{i}{g} \left[  \Big(V_{\xT}^{A}\partial^i V_{\xT}^{A~\dagger} \Big) \right.  \nonumber \\
&& \qquad \left. + \Big(V_{\xT}^{B}\partial^i V_{\xT}^{B~\dagger}  \Big) \right]\;, \label{eq:ini1} \\
&&E^{\eta}_{\xT}(\tau=0^{+})=\frac{i}{g} \left[  \Big(V_{\xT}^{A}\partial^i V_{\xT}^{A~\dagger}  \Big), \right. \nonumber \\ 
&& \qquad \left. \Big(V_{\xT}^{B}\partial^i V_{\xT}^{B~\dagger}  \Big) \right]\label{eq:ini2}\;.   
\end{eqnarray}
Here $A$ and $B$ refer to the two colliding nuclei, and the Wilson lines implicitly depend on evolution rapidity. As we consider midrapidity kinematics, the Wilson lines for both nuclei evolved to the same rapidity.
Discretized lattice representation of the initial conditions~\eqref{eq:ini1} and~\eqref{eq:ini2}
is used as an input for the classical Yang-Mills (CYM) equations of motion solved on
the lattice up to a time of $ \tau = 0.2 \, \text{fm}/c $~\cite{Krasnitz:1998ns,Schenke:2015aqa}.  At this point, we evaluate the energy-momentum tensor $ T^{\mu\nu} $, and obtain the energy density as
\begin{align}\label{eq:Ttautau}
    T^{\tau\tau} &{}= \frac{1}{4} e^{2y} (T^{00} + 2 T^{0z} + T^{zz}) + \frac{1}{2} (T^{00} - T^{zz}) \notag \\  &{}+ \frac{1}{4} e^{-2y} (T^{00} - 2 T^{0z} + T^{zz}).
\end{align}
Here $y=0$ corresponds to the midrapidity kinematics.

We consider collisions at zero impact parameter $b=0$ (corresponding to most central collisions) to minimize geometric asymmetries, enabling intrinsic nuclear deformation to dominate the resulting observables. We, however, perform random 3d rotations for the nuclei before the collision, as experimentally it is not possible to align the colliding nuclei (see however Ref.~\cite{Mantysaari:2024xmy} in the context of polarized light nuclei).

We calculate eccentricities at fixed Bjorken-$x$, corresponding to midrapidity kinematics and particle production at fixed transverse momentum. 
This is in contrast to IP-Glasma, where one typically uses a dynamically determined $\bt$-dependent $x$ defined as $x=Q_s^2(x,\bt)$~\cite{Schenke:2020mbo}. If the Bjorken-$x$ depends on the thickness function, it effectively also has a small effect on the apparent geometry. We choose to evaluate Wilson lines at fixed $x$ (using the \texttt{useFluctuatingx 0} option in IP-Glasma~\cite{ipglasma_code}), as it would be computationally very demanding to have $x$ dependent Wilson lines available. Furthermore, the fixed-$x$ setup allows us to more directly quantify the effects of the JIMWLK evolution on the nuclear shape. The difference between the two setups is formally higher order in $\as$, and as the evolution is only logarithmic in $x$, only moderate effects are expected. However, we note that the density profiles extracted in Sec.~\ref{sec:results} can be described by the Woods-Saxon distribution~\eqref{eq:WS} only when the Wilson lines are evaluated at fixed Bjorken-$x$.

\section{Results}
\label{sec:results}

We calculate the forward elastic dipole-nucleus scattering amplitude \eqref{eq:dipoleamp} for a fixed size  $|\rt|=0.1$ fm dipole 
as a function of impact parameter $b$ at different azimuthal angles $\theta$. As per the optical theorem, this is directly proportional to the cross-section for the fixed-size dipole-target scattering.
In the case of a small dipole considered in this work, the dipole amplitude is not highly sensitive to saturation effects and can be considered to be directly proportional to the local transverse density. This corresponds to assuming $T_A(\bt)\sim Q_s^2(\bt)$ consistently with Eq.~\eqref{eq:mv_rhorho}, which is also a typical estimate used to model the impact parameter dependence of the nuclear high-energy structure, see e.g. Refs.~\cite{Kowalski:2003hm,Lappi:2013zma}.
At each $\theta$, we fit a Woods-Saxon distribution~\eqref{eq:TA} (with a necessary normalization factor) to the impact parameter dependence of the scattering amplitude to obtain the effective radius $R(\theta)$. Finally, the deformation parameters $\beta_i$ are obtained by fitting the decomposition~\eqref{eq:deform_R} to the extracted radii.

\subsection{Uranium}

\begin{figure*}
    \centering
    \subfloat[$Y=0$]{
        \includegraphics[width=0.48\linewidth]{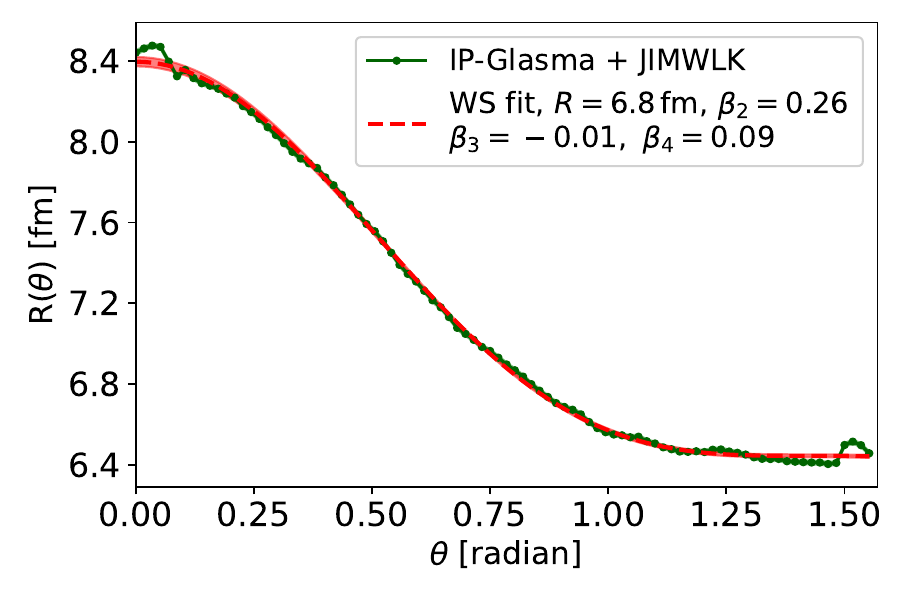}
        \label{fig:U_Y0r0.1}
    }
    \hfill
    \subfloat[$Y=4.8$]{
        \includegraphics[width=0.48\linewidth]{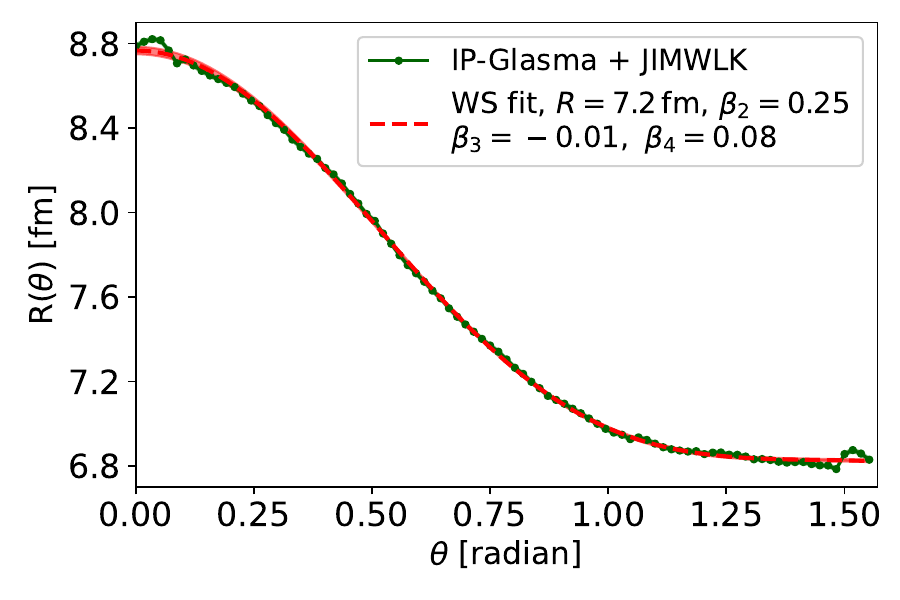}
        \label{fig:U_Y4.8r0.1}
    }
    \caption{Angle-dependent radius for the Uranium at the initial condition $Y=0$ and after $Y=4.8$ units of rapidity evolution. The radius parametrized as Eq.~\eqref{eq:deform_R} fitted to this data is also shown. 
    }
    \label{fig:U_radius_theta}
\end{figure*}

We begin our analysis by considering the ${}^{238}$U nucleus, which has a well-understood low-energy structure with a highly deformed geometry~\cite{Bemis:1973zza,Zumbro:1984zz}, and is also the heaviest nucleus used in collider experiments.
We construct Uranium nuclei using the Woods-Saxon parameters $R_\mathrm{WS}=6.81$ fm and $a=0.55$ fm, with the deformated structure described by $\beta_{2,\mathrm{WS}}=0.28, \beta_{3,\mathrm{WS}}=0$ and $\beta_{4,\mathrm{WS}}=0.093$ from Refs.~\cite{Bemis:1973zza,Zumbro:1984zz}. 
Here the subscript $\mathrm{WS}$ refers to the fact that these parameters enter the Woods-Saxon distribution~\eqref{eq:WS}.
As discussed in Refs.~\cite{Shou:2014eya,Ryssens:2018dza}, the multipole moments $\beta_{lm,\mathrm{WS}}$ of the Wood-Saxon density (in this work we only consider $m=0$) do not exactly correspond to the deformed geometry of the constructed nucleus $\beta_{lm}$, as the former represents the entire nuclear volume, while the latter describes surface deformation. Additionally, the shape parameters $\beta_{lm,\mathrm{WS}}$ can mix at higher orders, meaning that, for example, $\beta_{4,\mathrm{WS}}$ can contribute to the generation of $\beta_{2}$, reflecting the interplay of harmonic components.

We perform a JIMWLK evolution over $Y=4.8$ units of rapidity, from $x=0.01$ to $x=0.01 \cdot e^{-4.8} \approx 8\cdot 10^{-5}$, which covers the energy range from RHIC to LHC kinematics.
Although there are currently no plans to perform U+U collisions at the LHC,  this analysis allows us to quantify the importance of the geometry evolution between the two energy scales for any heavy nucleus.
Furthermore, we emphasize that the small-$x$ evolution over a comparable range is also probed by the rapidity dependence of e.g. flow measurements. This is because in such processes one probes the nuclear structure at $x\sim \langle p_T\rangle/\sqrt{s} e^{\pm y}$, where $\langle p_T \rangle$ is the typical transverse momentum of the parton produced at rapidity $y$. 

The effective nuclear radius as a function of angle $\theta$ (with respect to the axis aligned along the long axis of $\mathrm{U}$) is shown in Fig.~\ref{fig:U_radius_theta}. The results are shown both at the initial condition  ($Y=0$), and at the maximum evolution rapidity after $Y=4.8$ units of rapidity evolution. The nuclear radius is seen to grow towards small-$x$. When moving from RHIC to LHC energies, this growth is approximately $5\%$. This is comparable to the growth obtained in Ref.~\cite{Mantysaari:2022sux} and required to describe the ALICE data for the exclusive $\mathrm{J}/\psi$ photoproduction in ultra peripheral collisions~\cite{ALICE:2021tyx}.

We fit the function \eqref{eq:deform_R} to the calculated radii and extract the deformation parameters $\beta_n$. This fit is also shown in Fig.~\ref{fig:U_radius_theta} for comparison. The fit quality is found the be excellent at all rapidities. We choose to include coefficients up to $\beta_4$. Although lower harmonics $\beta_{n,\mathrm{WS}}$ can mix and in principle generate $n>4$ harmonics in the density profile, these higher-order coefficients turn out to be so small that they can not be reliably extracted numerically. In this work we choose to include the minimum number of deformation parameters required to obtain a good and stable fit to the calculated $R(\theta)$ data. The fit stability is determined by requiring that the obtained $\beta_i$ coefficients are not strongly modified if the fit is performed over a limited angle of $\theta$.

\begin{figure*}[tb]
\subfloat[$n=2$]{
\label{fig:U_b2}
\includegraphics[width=0.47\textwidth]{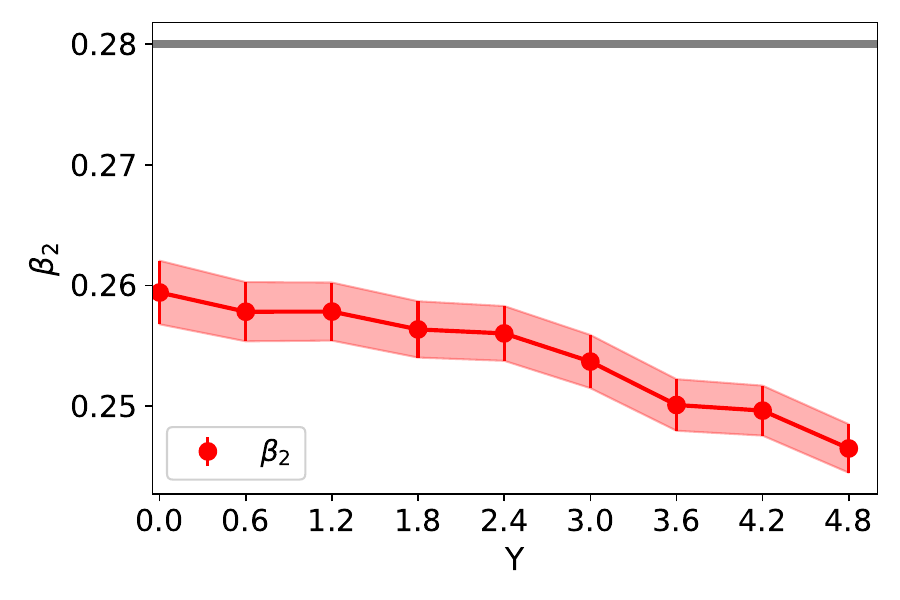}
}
\subfloat[$n=4$]
{
\label{fig:U_b3}
\includegraphics[width=0.47\textwidth]{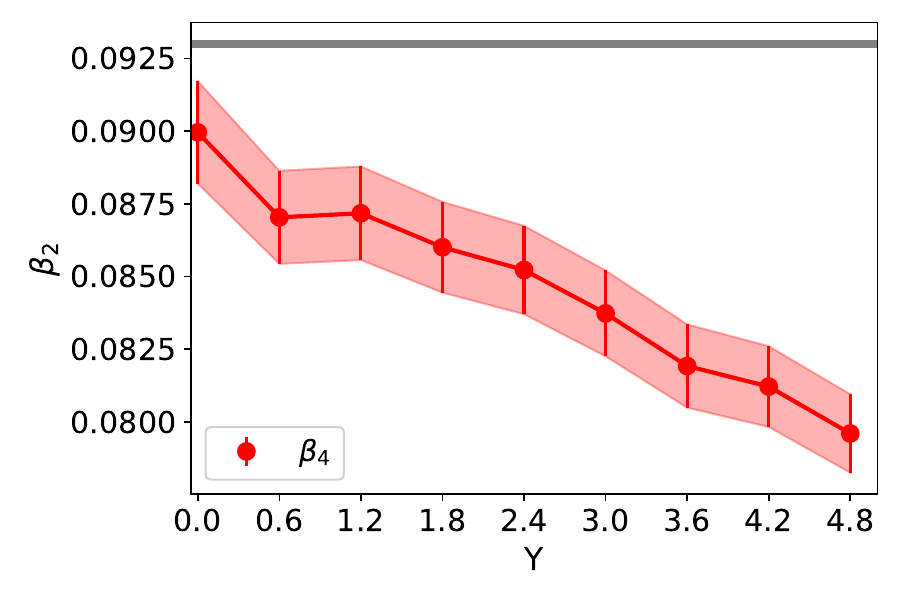}
}
    \caption{The obtained values of $\beta_2$ (left) and $\beta_4$ (right) deformation parameters as a function of rapidity for the Uranium. The grey line shows the corresponding value at the initial condition in the Woods-Saxon distribution. The result for $\beta_3$ (not shown) is small and approximatively compatible with zero.}
    \label{fig:U_beta_n_Ydep}
\end{figure*}

We extract the deformation parameters $\beta_n$ separately at each evolution rapidity $Y$. 
The rapidity dependence of the extracted parameters is shown in Fig.~\ref{fig:U_beta_n_Ydep}. The JIMWLK evolution is found to result in a smoother nucleus with increasing evolution rapidity (decreasing $x$), which manifests itself as $\beta_2$ and $\beta_4$ approaching zero. The $\beta_3$ is approximatively compatible with zero at all rapidities, i.e. it is not generated by non-zero $\beta_{2,4,\mathrm{WS}}$ at the initial condition or by the JIMWLK evolution. 
This geometry evolution is, however, only modest in the studied evolution range. For example, $\beta_2$ decreases only by $\sim 1.3\%$ from RHIC to LHC energies. 
For the $\beta_4$, which is sensitive to shorter distance-scale structure, the evolution is somewhat faster, with $\beta_4$ decreasing by $\sim 10\%$ over the same energy range.
This weak evolution is compatible with the previous studies of Uranium, Neon and Oxygen nuclei in Refs.~\cite{Mantysaari:2023qsq,Singh:2023rkg}, showing that the non-trivial shape is not significantly altered by the small-$x$ evolution. 
On the other hand, in 
Ref.~\cite{Mantysaari:2023xcu} it has been shown that at the nucleon substructure level, the nucleus becomes much smoother as a result of JIMLWK evolution, which has a numerically large effect on the 
 incoherent $\mathrm{J}/\psi$ production in $\gamma+\mathrm{Pb}$ scattering. 
 Similarly a significant geometry evolution for individual protons with non-trivial initial geometry has been reported in Refs.~\cite{Schlichting:2014ipa, Schenke:2022mjv,Mantysaari:2018zdd}.
 This suggests that the small-$x$  evolution has  a moderate effect on the nuclear geometry at the largest length scales, but can have a more pronounced effect at the length scales comparable to and shorter than the nucleon radius.

Next, we determine the relation between the deformed geometry of the Uranium nucleus and the initial state eccentricities in U+U collisions. As discussed in Sec.~\ref{sec:setup}, we consider U+U collisions at central impact parameter $b=0$, but perform random 3d rotations for both nuclei. As such, this in principle corresponds to actual heavy ion collisions that can be studied in collider experiments (although in U+U collisions the relationship between the impact parameter and different centrality estimators can be complex~\cite{Goldschmidt:2015kpa}), unlike the dipole-nucleus scattering amplitude studied above.

\begin{figure*}[tb]
    \centering
    \includegraphics[width=0.45\linewidth]{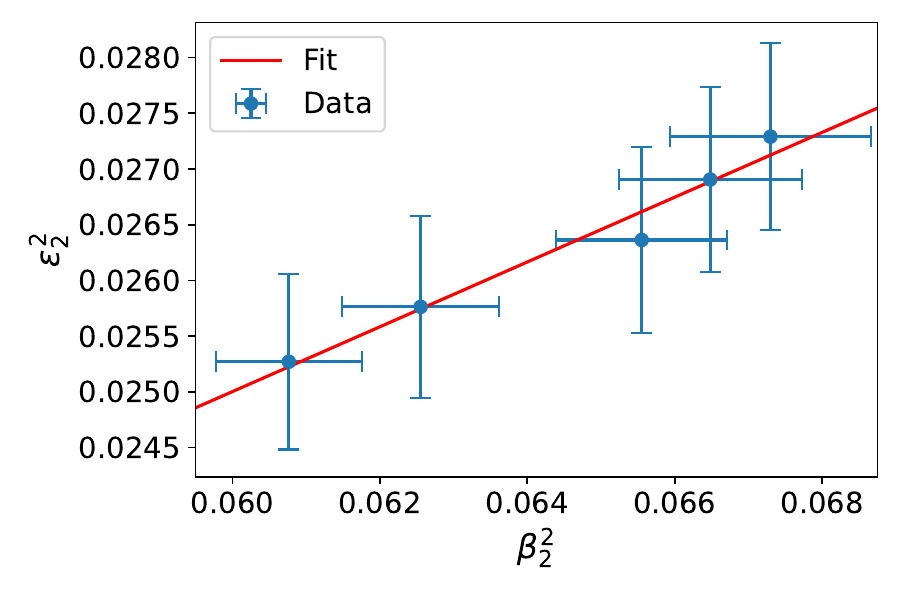}
    \includegraphics[width=0.45\linewidth]{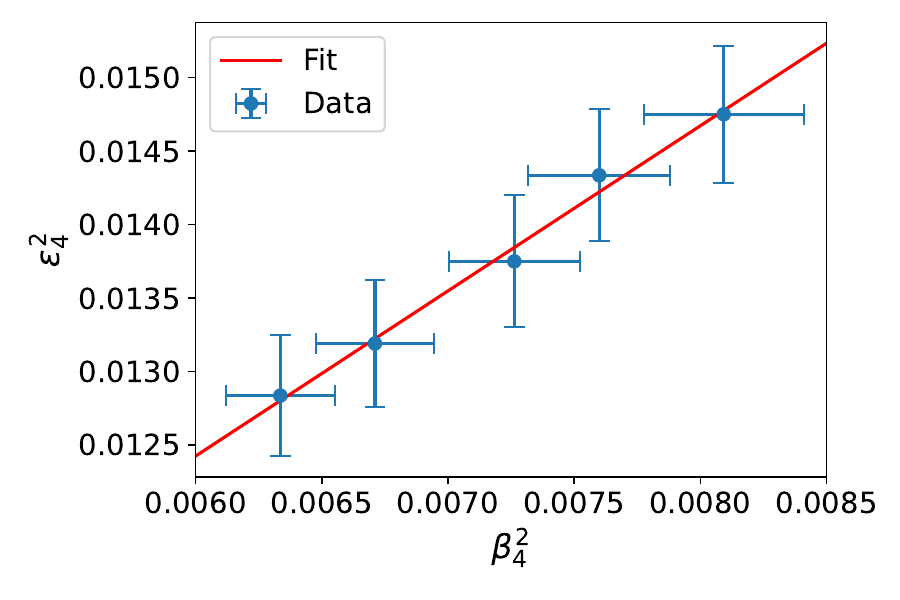}
    \caption{Eccentricity as a function of nuclear deformation parameter. The line shows the best fit of the function~\eqref{eq:rel_ecc_b} }
    \label{fig:eccentricity_UU}
\end{figure*}

The correlations between the squared eccentricities $\varepsilon_n^2$ and deformation parameters $\beta_n^2$ are shown in Fig.~\ref{fig:eccentricity_UU} for the second and fourth order coefficients. For the third-order coefficient, the results are compatible with zero. Interestingly, although eccentricities correspond to the case where the Uranium nuclei have random orientations, the JIMWLK evolution has numerically a comparable effect on both eccentricities and on deformation parameters.

We find the linear relation suggested in Ref.~\cite{Zhang:2021kxj}, 
\begin{equation}
\label{eq:rel_ecc_b}
    \varepsilon_n^2 = a_n+b_n\beta_n^2
\end{equation}
to be very accurate in central collisions. The fitted coefficients $a$ and $b$ are shown in Table.~\ref{table:fit_coefs_U}. The $a_n$ coefficients are found to be approximatively zero as expected, as the system becomes rotationally symmetric in the $\beta_n\to 0$ limit.  The correlation between $\beta_4$ and $\varepsilon_4$ is found to be stronger than in the $n=2$ case. These correlation results demonstrate that the Bjorken-$x$ dependence of the deformed nuclear geometry is visible at the level of eccentricities, and as such expected to influence precise flow measurements. This also suggests that the rapidity dependence of the flow measurements can be sensitive to the small-$x$ evolution.

\begin{table}[tb]
\centering
\begin{tabular}{|c|c|c|}
    \hline
     &  $n = 2$ & $n = 4$ \\
    \hline
    ~~~~~a~~~~~ & ~~~~~ $0.008 \pm 0.009$~~~~~ & ~~~~~ $0.006 \pm 0.002$~~~~~  \\
    \hline
    ~~~~~b~~~~~ & $0.291 \pm 0.147$ & $1.124 \pm 0.316$ \\ 
    \hline
\end{tabular}
\caption{Fitted coefficients quantifying the correlation between the squared eccentricities and deformation parameters as defined in Eq.~\eqref{eq:rel_ecc_b}.
}

\label{table:fit_coefs_U}
\end{table}

\subsection{Ruthenium}

In addition to the heaviest nuclei, there is also a growing interest to probe intermediate-$A$ nuclei~\cite{Brewer:2021kiv,Bally:2022vgo}. 
In order to study geometry evolution in the case of intermediate nuclei, we consider Ruthenium in this work. One additional motivation for this study is the fact that the JIMWLK evolution can be expected to have a pronounced effect on the geometry of smaller light nuclei. Additionally, on a phenomenological level, a precise knowledge of the deformed structure of $^{96}\text{Ru}$ and $^{96}\text{Zr}$ is essential due to their involvement in high-energy collisions at RHIC, aimed at probing local strong parity violation. Deviations from unity in the ratio of observables for $^{96}\text{Zr}+^{96}\text{Zr}$ and $^{96}\text{Ru}+^{96}\text{Ru}$ collisions are predominantly attributed to differences in their radial profiles and intrinsic deformations~\cite{STAR:2021mii}. Precise characterization of these deformations is crucial for interpreting experimental data and isolating potential parity-violating effects.

\begin{figure*}[tb]
    \centering
    \subfloat[$Y=0$]{
    \includegraphics[width=0.47\linewidth]{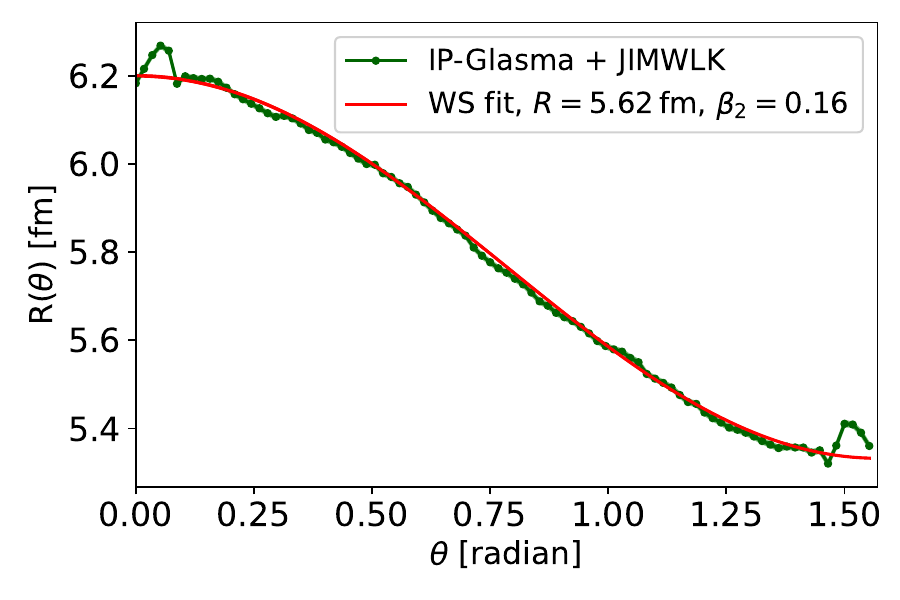}
    }
    \subfloat[$Y=4.8$]{
    \includegraphics[width=0.47\linewidth]{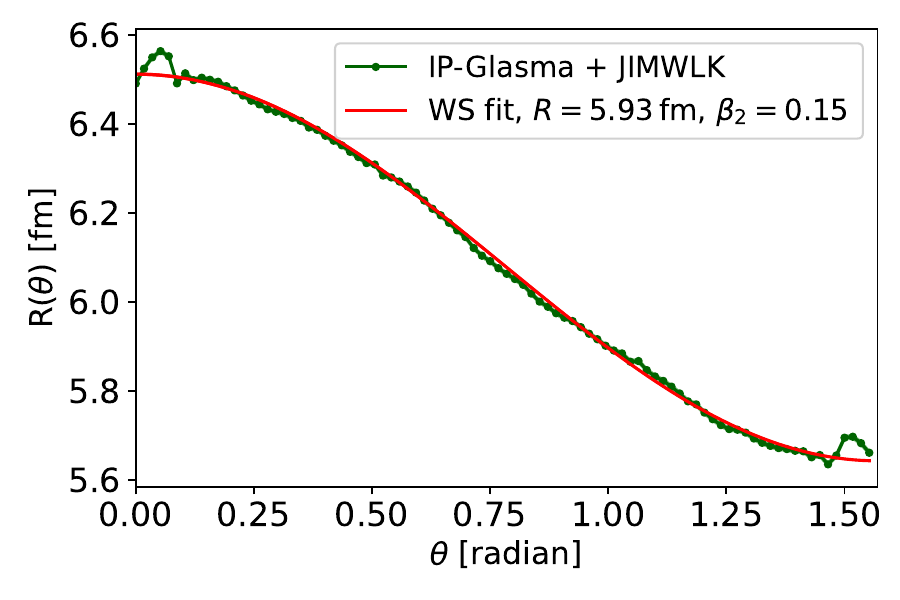}
    }
    \caption{Angle-dependent radius for the Ruthenium at the initial condition Y = 0 and after Y = 4.8 units of rapidity evolution. 
    }
    \label{fig:Ru_radius_theta}
\end{figure*}
There is only quadrupole deformation in Ruthenium, i.e. $\beta_{2,\mathrm{WS}}=0.158$, and $\beta_{3,\mathrm{WS}}=\beta_{4,\mathrm{WS}}=0$ \cite{Shou:2014eya}. The other parameters for the Woods-Saxon distribution are $R=5.085$ fm and $a=0.46$ fm.
We only achieve a stable fit to the $R(\theta)$ distribution when including the $\beta_2$ coefficient and setting the higher-order coefficients to zero. An excellent description of the $R(\theta)$ distribution with only one free deformation parameter $\beta_2$ is illustrated in Fig.~\ref{fig:Ru_radius_theta}. Consequently, we only extract $\beta_2$ for Ruthenium and conclude that the other deformation parameters are negligible. We note that although there are large fluctuations in the calculated $R(\theta)$ data close to $\theta=0$ and $\theta=\pi/2$, the obtained $\beta_2$ or radius $R$ are not significantly affected if that part is not included in the fit.

\begin{figure}[tb]
    \centering
    \includegraphics[width=\columnwidth]{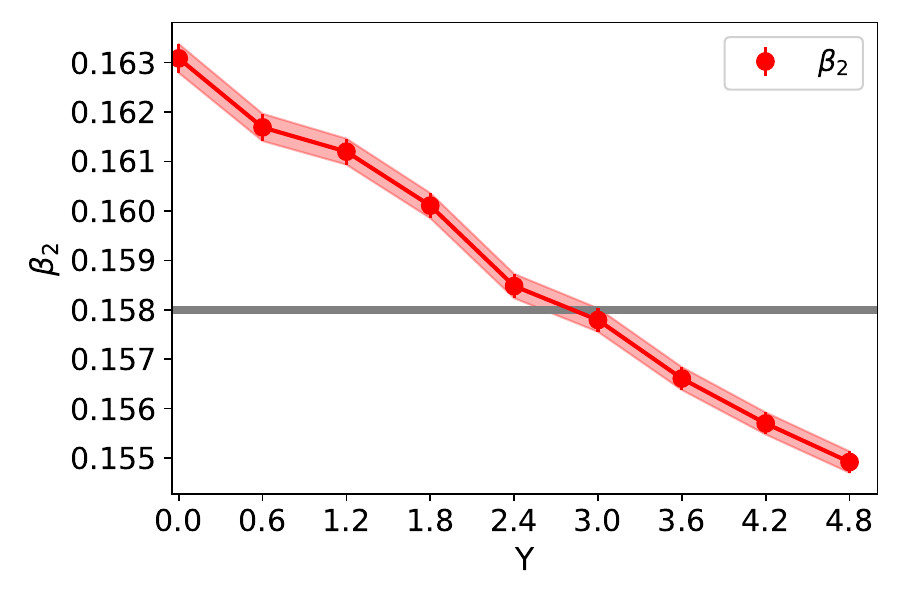}
    \caption{The values of $\beta_2$ as a function of rapidity, derived from the fit to data using the Ruthenium setup.}
    \label{fig:Ru_beta_ydep}
\end{figure}

Similarly, as in the case of the Uranium, the Ruthenium radius increases by approximately 5\% when evolving from RHIC to LHC energies. The rapidity dependence of the $\beta_2$ deformation parameter is shown in Fig.~\ref{fig:Ru_beta_ydep}. The JIMWLK evolution is again found to drive the nuclear geometry more towards a spherical shape. The $\beta_2$ coefficient now decreases by $\sim 5\%$ in the considered energy range, which is a much larger effect than what was seen in the case of Uranium. This stronger $x$ dependence in $\beta_2$ we attribute to the smaller size of the Ruthenium nucleus which is more effectively modified by the emission of soft gluons throughout the JIMWLK evolution.

\begin{figure}
    \centering
    \includegraphics[width=\linewidth]{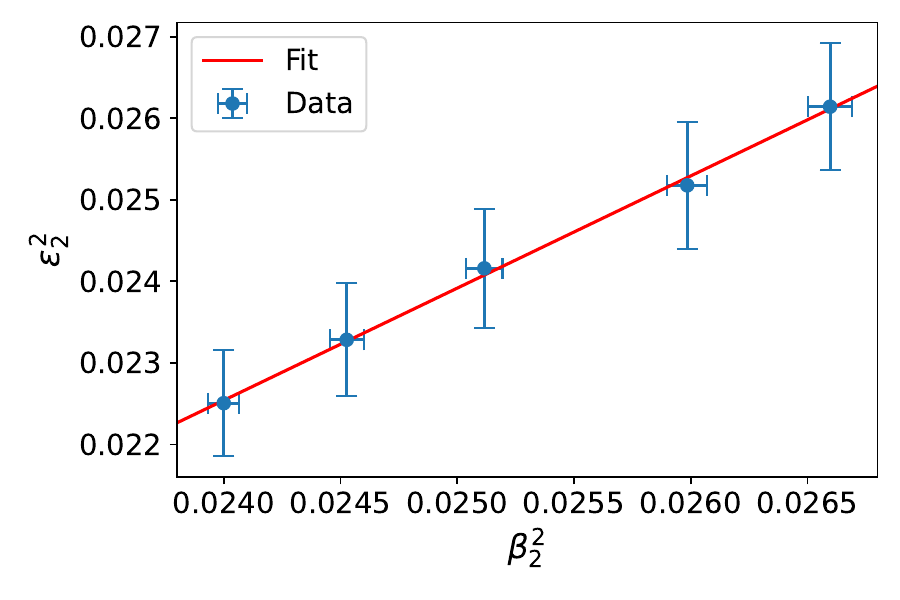}
    \caption{Eccentricity as a function of the nuclear deformation parameter. The line represents the best fit to the Ruthenium data}
    \label{fig:Ru_beta_e2}
\end{figure}

The correlation between the quadrupole deformation $\beta_2$ and the eccentricity $\varepsilon_2$ is shown in Fig.~\ref{fig:Ru_beta_e2}. Again we perform random rotations for the Ru nuclei before the collision and consider only the central $b=0$ events. Similarly, as in the case of the Uranium, we find a perfect linear correlation between $\varepsilon_2^2$ and $\beta_2^2$, with the coefficients shown in Table~\ref{table:fit_coefs_Ru}.  This confirms that the linear relation \eqref{eq:rel_ecc_b} is valid for both intermediate and heavy nuclei with deformed geometries in central collisions. Compared to Uranium, a stronger correlation between $\beta_2$ and $\varepsilon_2$ is found.  This again suggests that a precise description of the $x$ dependent geometry is crucial when confronting simulations with precise collider data.

\begin{table}
\centering
\begin{tabular}{|c|c|c|}
    \hline
     &  $n = 2$  \\
    \hline
    ~~~a~~~ & ~~~ $-0.01 \pm 0.008$~~~   \\
    \hline
    ~~~b~~~ & $1.376\pm 0.343$  \\ 
    \hline
\end{tabular}
\caption{Coefficients for Ruthenium for  $n = 2$ }
\label{table:fit_coefs_Ru}
\end{table}

\section{Conclusions}

We have calculated the small-$x$ evolution of the deformed nuclear geometry, focusing on both heavy (Uranium) and intermediate (Ruthenium) nuclei. The energy dependence of the nuclear geometry is obtained by numerically solving the  JIMWLK evolution equation. 

The JIMWLK evolution is found to drive the nuclear shape towards a larger and more rotationally symmetric form, decreasing the deformation parameters $\beta_n$ as defined in Eq.~\eqref{eq:deform_R}. This evolution is, however, found to be rather slow, having only a few percent effect for the dominant quadrupole deformation parameter $\beta_2$ for intermediate and heavy nuclei, when evolving from RHIC to LHC energies. The evolution is stronger in the case of smaller nuclei or higher-order deformations. 
This can be understood to result from the fact that the confinement scale effects effectively included in the JIMWLK kernel limit the relevant length scales in the evolution to not significantly exceed the nucleon radius. 

In this work, a fixed coupling evolution is used, and consequently,y our results can be seen to correspond to a lower limit for the effect of the JIMWLK evolution on the deformed geometry. This is because including running coupling effects would enhance the evolution at long-distance scales relevant for deformation, see also discussion in Ref.~\cite{Mantysaari:2018zdd}.

We furthermore demonstrated that, for both Uranium and Ruthenium, the linear relation $\varepsilon_n^2 = a + b \beta_n^2$ connecting the eccentricities and deformation parameters, is valid in central collisions. This demonstrates that the deformed energy-dependent geometry leaves visible signatures in central nucleus-nucleus collisions even when nuclei undergo a random rotation before the collision as in an actual experiment. Given the high precision of the flow measurements performed at RHIC and at the LHC, a precise understanding of the energy-dependent geometry is crucial to understand the accurate flow data. In addition to the center-of-mass energy dependence, comparable signatures of the JIMWLK evolution in the nuclear geometry can be expected to be visible in rapidity-dependent flow observables. In the future, it would be interesting to both include the running coupling effects and to determine the signatures of the JIMWLK evolved geometry at the level of actual flow measurements in e.g. experimentally accessible multiplicity bins.

\begin{acknowledgements}
This work was supported by the Academy of Finland, the Centre of Excellence in Quark Matter (project 346324), and projects 338263,  346567 and 359902, and by the European Research Council (ERC, grant agreements No. ERC-2023-101123801 GlueSatLight and No. ERC-2018-ADG-835105 YoctoLHC). The content of this article does not reflect the official opinion of the European Union and responsibility for the information and views expressed therein lies entirely with the authors. Computing resources from CSC – IT Center for Science in Espoo, Finland and from the Finnish Computing Competence Infrastructure  (persistent identifier \texttt{urn:nbn:fi:research-infras-2016072533}) were used in this work.  
\end{acknowledgements}

\bibliography{refs}
\bibliographystyle{JHEP-2modlong}

\end{document}